\documentclass[twocolumn,pre,amsmath,amssymb,showpacs,aps]{revtex4-1}

\usepackage{graphicx}
\usepackage{color}
\usepackage{epstopdf}
\begin{document}

\preprint{APS/123-QED}

\title{Analytically solvable processes on networks}

\author{Daniel Smilkov} 
\affiliation{Macedonian Academy for Sciences and Arts, Skopje, Macedonia \\ 
Email: dsmilkov@cs.manu.edu.mk}

\author{Ljupco Kocarev}
\affiliation{Macedonian Academy for Sciences and Arts, Skopje, Macedonia \\
 BioCircuits Institute, University of California, San Diego \\
 9500 Gilman Drive, La Jolla, CA 92093-0402 \\ Email: lkocarev@ucsd.edu}

\begin{abstract}

We introduce a broad class of analytically solvable processes on networks. In the special case, they reduce to random walk and consensus process -- two most basic processes on networks. Our class differs from previous models of interactions (such as stochastic Ising model, cellular automata, infinite particle system, and voter model) in several ways, two most important being: (i) the model is analytically solvable even when the dynamical equation for each node may be different and the network may have an arbitrary finite graph and influence structure; and (ii) in addition, when local dynamic is described by the same evolution equation, the model  is decomposable: the equilibrium behavior of the system can be expressed as an explicit function of network topology and node dynamics. 

\end{abstract}


\maketitle

\section{Introduction} 

Network structures are pervasive throughout biological, information, social, and technical systems. 
Recently network theory has paved the way for exploring many real-world large-scale networks, and describing and understanding various processes  taking place on these networks  \cite{ref-0}. Examples of such processes include virus propagation in social and computer networks; the diffusion of innovations, opinion formation, and search processes in social networks; routing packets in communication networks, to name just a few. 
Two most fundamental processes on networks are random walk and consensus process. Random walks on networks are dynamical processes aiming at modeling the diffusion of some quantity or information on networks. 
They can be used to model random processes inherent to many important applications, such as transport in disordered media \cite{ref-1}, neuron firing dynamics \cite{ref-2}, spreading of diseases \cite{ref-3}, or transport and search processes \cite{ref-4,ref-5,ref-6,ref-7,ref-8}.
Another kind of processes on networks, related to random walks, are so-called ``distributed consensus processes''. In networks of dynamic systems (or agents), consensus means to reach an agreement regarding a certain quantity of interest that depends on the state of all dynamical systems (agents). Consensus problems have a long history in computer science and form the foundation of the field of distributed computing \cite{ref-9,ref-10,ref-11,ref-12,ref-13,ref-14}.

In this paper we introduce a broad class of analytically solvable processes on networks. We argue that this class can model various interactions on networks as well as hierarchical organization of complex systems (networks). An example of a process belonging to this class is a network of Markov chains: dynamics of each node is governed with a Markov chain having arbitrary number of states and transition probabilities that depend on, not only the current states of that node, but also on the states of the neighboring nodes. Another example is a network of (heterogeneous) agents in which agreement between states of the agents should be reached. Yet another example is generalized random walk where walkers move between not only the nodes but also between the internal states of all the nodes.
The class differs from previous models of interactions, such as stochastic Ising model \cite{binder}, cellular automata \cite{cel-auto}, infinite particle system \cite{infinite-ps}, and voter model \cite{voting}. 
We show that for homogeneous case, when each node (local) dynamic is described by the same evolution equation, the solution of the model is decomposable in the sense that it depends on the eigenvectors of two matrices: a matrix related to the adjacency matrix of the network and a matrix describing the local dynamics. 
Moreover, it is shown that the heterogeneous model is analytically solvable even when the dynamical equation for each node may be different and the network may have an arbitrary finite graph and influence structure.

The outline of this paper is as follows. In Section \ref{sec:examples} we introduce a broad class of linear processes on networks which we refer to as general linear processes. To present the ubiquity of these processes we include several real-world examples. Then, we split these processes in two main classes based on the nature of local dynamics and discuss each class in seperate sections. Thus, in Section \ref{sec:homo} we look more closely into the homogeneous processes where the local dynamics in each node is described with the same evolution equation and discuss three variants of this class. Then, in Section \ref{sec:hetero} we focus on the second class which we call heterogeneous processes where the local dynamics differs for each node in the network and again discuss three variants of this class of processes. 
Section \ref{sec-net-hier} shows how linear processes on networks can be used to model and/or to understand  interactions on hierarchical complex systems.
Lastly, in Section \ref{sec:concl} we conclude our paper.

\section{Linear processes on networks} \label{sec:examples}
In the following, we focus on undirected, connected and non-bipartite networks, which are described by their $N \times N$ symmetric adjacency matrix $A=[a_{ij}]$, where $N$ is the number of nodes. By definition, $a_{ij}$ is the topological weight of the edge $ij$. The strength $s_i = \sum_j a_{ij}$ of node $i$ is the total weight of the links connected to it. If the network is unweighted, $s_i$ is simply the degree of node $i$. $W = \sum_{ij} a_{ij}/2$ is the total weight in the network.

The simplest dynamical processes on networks are linear processes:
\begin{equation}
x_i(t+1) =\sum_j
b_{ij}x_j(t), \label{eq-linear} 
\end{equation}
where the evolution of a quantity $x_i$, associated to node $i$, is driven by $B=[b_{ij}]$, a matrix related to the adjacency matrix $A$. Here we focus on two related linear processes: unbiased random walk and consensus. Both processes can be written in more compact form as:
\begin{equation}
\label{eq-linear-matrix} 
\mathbf{x}(t+1)=B\mathbf{x}(t),
\end{equation}
where  $\mathbf{x}=[x_1, \ldots, x_N]^T$ is column vector of length $N$. The difference between the two processes is that for random walk $B$ is column stochastic, $b_{ij}=a_{ij}/s_j$, and for consensus $B$ is row stochastic, $b_{ij}=a_{ij}/s_i$. The processes converge to
\begin{eqnarray} \label{eq-lin-res} 
\mathbf{x}(t)=B^t \mathbf{x}(0) \rightarrow \begin{cases} \nonumber 
	(\pi \otimes \mathbf{1}_N^T)\mathbf{x}(0) & \text{random walk}\\
	(\pi^T \otimes \mathbf{1}_N)\mathbf{x}(0) & \text{consensus}
\end{cases} \\
 = \begin{cases}
	\pi \mathbf{1}_N^T\mathbf{x}(0)=\pi\left\| \mathbf{x}(0) \right\| \\
	\mathbf{1}_N \pi^T \mathbf{x}(0)
\end{cases}
\end{eqnarray}
where $\pi$ is the dominant eigenvector of $B$, $\mathbf{1}_N$ is a length $N$ column vector of 1, and $C \otimes D$ is a Kronecker product of matrices (or vectors) $C$ and $D$. 
We see that the solution of consensus depends on the initial vector $\mathbf{x}(0)$, and each node's opinion converges to a mixture of the initial opinions in the network. However, the random walk solution, which gives the fraction of time random walkers spent in each node, depends only on the number of concurrent random walkers in the graph, i.e. $\left\| \mathbf{x}(0) \right\|= \sum_{i=1}^N x_i(0) $. Also, the number of walkers is consistent in time, $\left\| \mathbf{x}(t+1) \right\|=\left\| \mathbf{x}(t) \right\|$, and it is usually 1, while in consensus processes, the aggregated opinion in the network changes with time and lacks consistency.

The main purpose of this work is to generalize the model (\ref{eq-linear-matrix}) and the result (\ref{eq-lin-res}). 
To do this, we assume that node $i$ is not described with a scalar quantity, rather a vector of quantities is associated to the node.  
We now consider two main approaches that take advantage of this generalization.
The first approach is to consider each node $i$ as a complex system with local behavior described with equation 
\begin{eqnarray} \label{eq01}
\mathbf{x}_i(t+1) &=& D_{ii} \mathbf{x}_i(t)
\end{eqnarray}
where $\mathbf{x}_i = [x_i^1\;x_i^2 \ldots x_i^{m_i}]^T $ is a nonnegative $m_i$-dimensional column vector and $D_{ii}$ is $m_i \times m_i$ stochastic matrix. Let as above $B$ be a stochastic $N\times N$ matrix related to the adjacency matrix $A$. We allow $b_{ii} \neq 0$. $b_{ij}$ contains information about the connection between nodes $i$ and $j$ in the network.
Then, for each pair of nodes $i$ and $j$, let $D_{ij}$ be an $m_i \times m_j$ nonnegative matrix such that each row (column) of $D_{ij}$ sums up to 1. Matrix $D_{ij}$ describes the dynamics between nodes $i$ and $j$.
Assume that the evolution of each node variables has the following linear form:
\begin{eqnarray} \label{eq1}
\mathbf{x}_i(t+1) &=&  \sum_{j=1}^Nb_{ij} D_{ij}\mathbf{x}_j(t),
\end{eqnarray}
for all $i=1, \ldots N$. 
The second approach is to consider a network with $N$ nodes, with each node $i$ actually being a network with $m_i$ internal nodes. Thus, the total number of nodes in this network of networks is $m_1 +\ldots +  m_N$.  In this case, $D_{ij}$ are $m_i \times m_j$ nonnegative matrices derived from the internal structure (topology) of the network $i$ as well as from the connections among the networks $i$ and $j$. Again, we consider the simplest dynamical processes on this network of networks which is a linear process described with Eq.~(\ref{eq1}).

In this paper we show that the model (\ref{eq1}) is simple enough so that it is analytically solvable, yet it is rich enough so it encounters various phenomena. In particular, when $m_i=1$ for all $i$, the model reduces to two basic processes on networks: random walk and consensus. 
The model (\ref{eq1}) consists of a network of nodes, each with states that evolve over time. The evolution of the states at a node depends on the current states of that node as well as on the states of the neighboring nodes.
The graph structure is described by the matrix $B$ and the influence structure is described by the matrix $B$ and the matrices $D_{ij}$. 
Note that when $b_{ii}=1$ for all $i$, (\ref{eq1}) reduces to $N$ unconnected systems (\ref{eq01}).
From now on, let $\mathbf{x} = [ \mathbf{x}_1 \ldots \mathbf{x}_N]^T$ be a column vector of length $m_1 +\ldots +  m_N$ and should not be confused with $\mathbf{x}$ from model (\ref{eq-linear}).

We now present several examples which also serve as our motivation for this work.

\begin{figure}[htb]
\begin{center}
\includegraphics[scale=0.6]{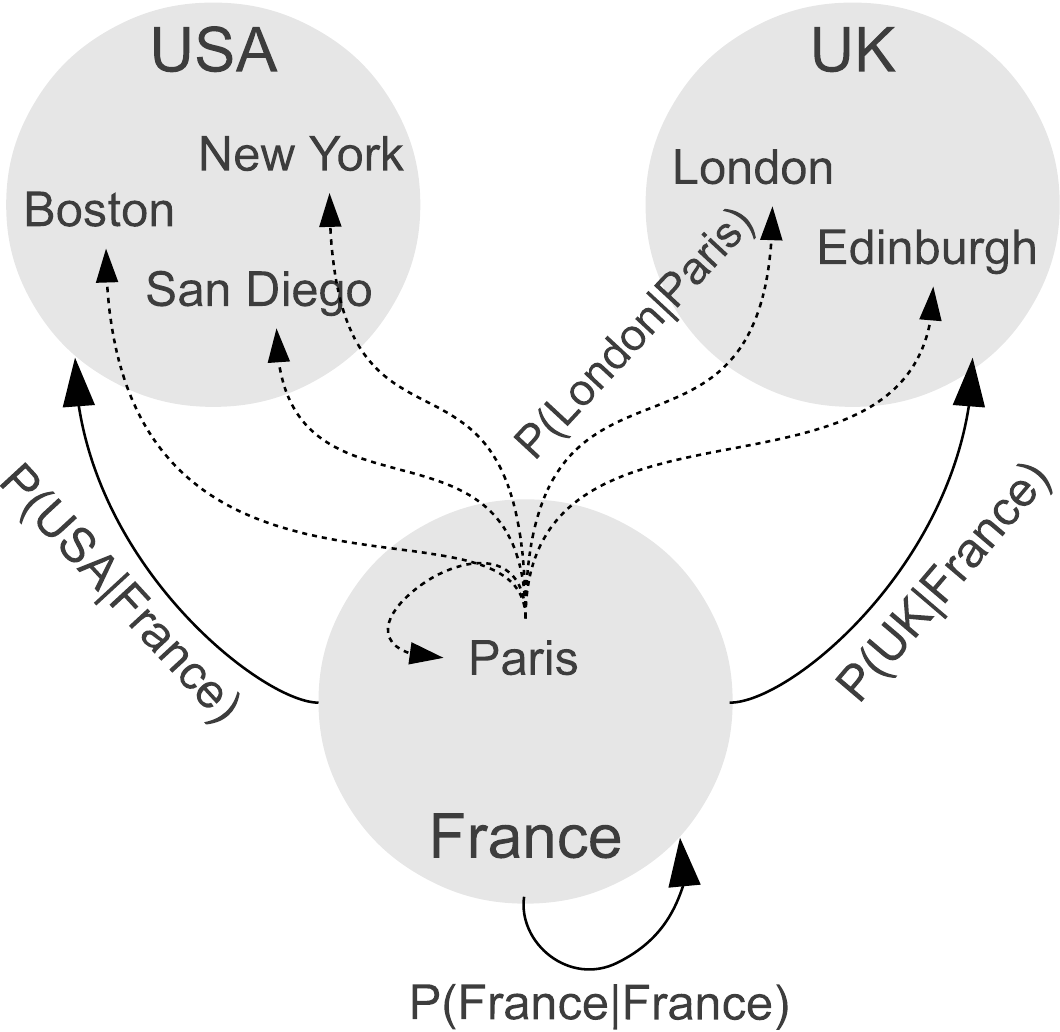}
\end{center}
\caption{ 2-step random walk where a walker makes a 2-step decision for where to go; it first chooses a country, then a city within that country. Dashed lines depict the probabilities for jumping from Paris to any other city and solid lines depict the transition probabilities from France to any other country. Labels on dashed lines (except one) are omitted for clarity.}
\label{fig:countries}
\end{figure}

\textbf{Example 1.}
In a network with $N$ nodes, assume that each node $i$ is actually a network with $m_i$ internal nodes. Thus, the total number of nodes in the network of networks is $m_1 +\ldots +  m_N$.
We are studying 2-step random walk where the walker makes a 2-step decision for where to go; first, it decides to jump to one of the external nodes having in mind the current external node, and then decides to jump to one of the internal nodes within the selected external node, this time having in mind the current internal node.
To illustrate this further, we depict an example in Fig. \ref{fig:countries} where the 2-step hierarchy consists of countries and cities within the countries respectively. A random walker travels across cities by first choosing a country having in mind the current country, then a city within the chosen country having in mind the current city he is in.
We denote with $b_{ij}$ the transition probability from country $j$ to country $i$ and with $d_{ij}^{kl}$ the transition probability from $l$-th city in country $j$ to $k$-th city in country $i$. $b_{ii}$ can be though of as the absorbing factor for country $i$, i.e. some countries can be more attractive than others and thus, more difficult for walkers to leave them. For convenience, we create matrices $B=[b_{ij}]$ and $D_{ij}=[d_{ij}^{kl}]$. Since the walker has to make a separate decision in each step, matrices $B$ and $D_{ij}$ for all $i,j$ are column stochastic. Then, encoding with vector $\mathbf{x}_i(t)=[x_i^1(t)\;x_i^2(t)\ldots x_i^{m_i}(t)]$ the expected density of random walkers in the $m_i$ cities in country $i$, the evolution of the random walker gets the form Eq.~(\ref{eq1}).


\begin{figure}[htb]
\begin{center}
\includegraphics[scale=0.6]{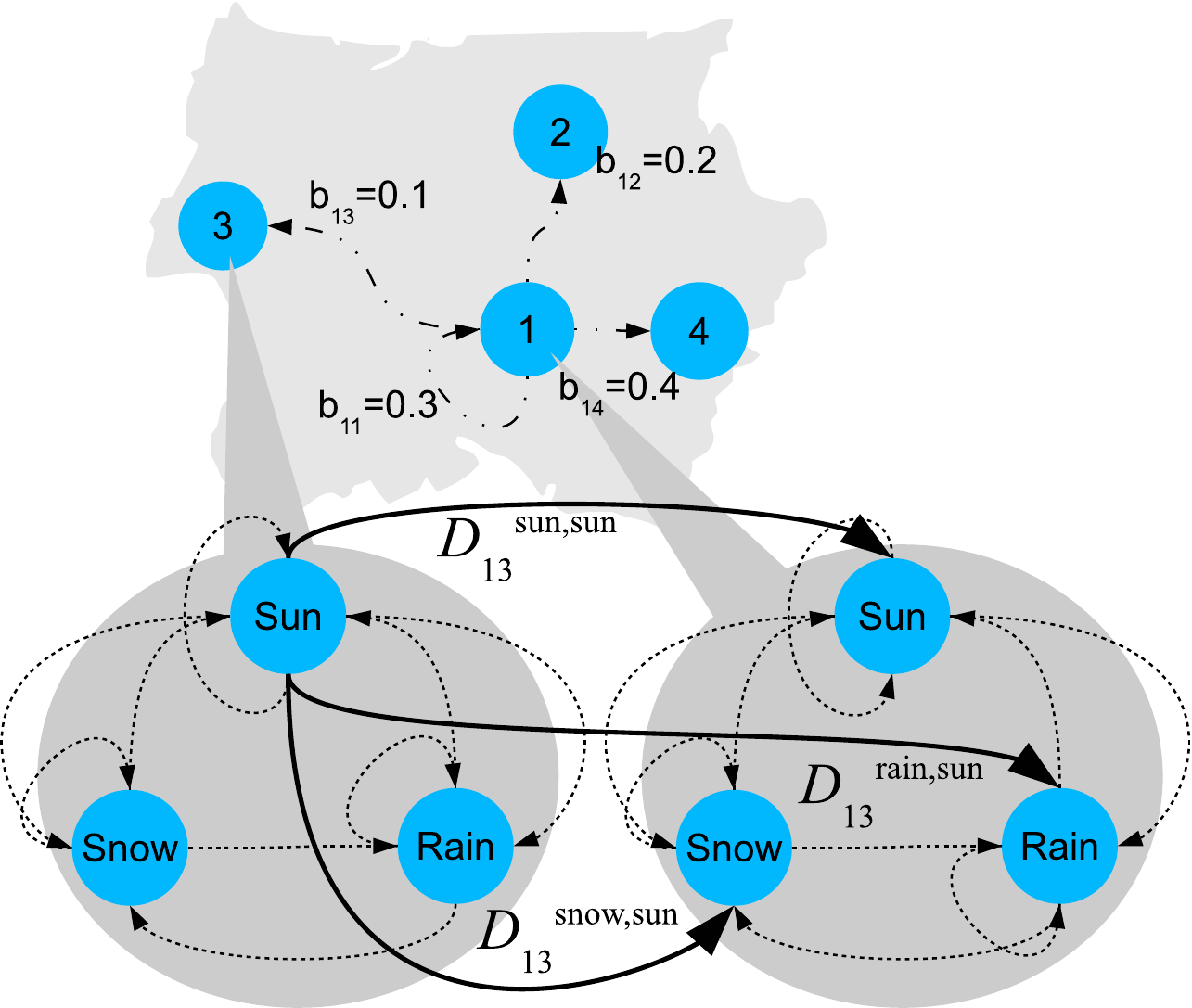}
\end{center}
\caption{ Network of Markov chains describing weather dynamics. Each node (physical location) corresponds to a Markov chain with 3 internal states: ``sun'', ``snow'' and ``rain''. Dashed lines depict the local transition probabilities. Solid lines depict inter-node transition probabilities (only those from ``sun'' state in node 3 are shown for clarity). Dash-dotted lines depict the influences neighboring nodes have on node 1 and are based on geographical proximity.}
\label{fig:weather}
\end{figure}

\textbf{Example 2.}
In a network with $N$ nodes, assume that each node $i$ corresponds to a Markov chain with $m_i$ states described by the vector $\mathbf{x}_i$. 
We are studying a network of Markov chains with transition probabilities that depend on, not only the current states of that node, but also on the states of the neighboring nodes.
To illustrate this further, we depict an example in Fig. \ref{fig:weather} where nodes correspond to physical locations, and each node can be in one of the 3 states describing the weather in that location : sun, rain and snow respectively. 
Therefore, each location $i$ represents an internal Markov chain with $m_i=m=3$ states. Let $D_{ii}$ be a column stochastic transition matrix for the internal Markov chain in location $i$. Thus, matrix $D_{ii}$ describes the local dynamics for location $i$.
Additionally, assume that the local weather can be affected by the weather in neighboring locations. Let $b_{ij}$ denote the relative influence location $j$ has on location $i$. This influence can be based on an arbitrary information such as proximity. Some locations have more stable weather that others, thus $b_{ii}$ can denote the stability of the weather in location $i$. The $r$-th row in matrix $B=[b_{ij}]$ contains the relative influences locations have on location $r$ and, unlike in example 1, here $B$ is row stochastic.
Furthermore, it is not uncommon to think for example that snowing in a neighboring location can increase the local probability of rain, and not just snow. To include this information, we add $3\times3$ ($m_i \times m_j$) column stochastic transition matrix $D_{ij}$ with $d_{ij}^{kl}$ denoting the transition probability from state $l$ in location $j$ to state $k$ in location $i$. Thus, matrices $D_{ij}$ together with matrix $B$ describes the global dynamics in the network.
The evolution of this system can also be described with Eq.~(\ref{eq1}). We stress again that this system differs from the one in example 1 since, here $B$ is row stochastic, where in example 1, $B$ is column stochastic.

\textbf{Example 3.}
Assume a model of opinion formation in a network of $N$ people where people discuss matters on $m_i=m$ topics and the opinion of person $i$ at time $t$ is denoted by $m$-dimensional vector $\mathbf{x}_i(t)$ where $x_i^k$ contains the opinion on $k$-th topic. People communicate and exchange their opinion with others. Since a person listens to (trusts) some people more than others, we use $N\times N$ matrix $B$ where $b_{ij}$ denotes the relative influence person $j$ has on person $i$. Here, $b_{ii}$ can describe the stubbornness (ressistance to other people's opinion) of person $i$. Consequently, the $r$-th row in matrix $B=[b_{ij}]$ contains the relative influences people have on person $r$, and thus, sums to 1. Each person's opinion is a combination of $m$ topics and interactions can be more complex where the opinion of one topic can be influenced not only by opinions in the same topic, but also opinions of different topics as well. To incorporate this, we introduce $m\times m$ matrix $D_{ij}=D$ which describes the dynamics of how opinions change after interaction. More specifically, let the opinion on $r$-th topic be a weighted average of the opinions of other topics with the weights encoded in the $r$-th row of matrix $D$. Please note that unlike the first two examples, here matrix $D$ (describing local dynamics) is row stochastic. This model can also be described with Eq.~(\ref{eq1}). 



\section{Homogeneous linear processes} \label{sec:homo}

We first consider the case: $m_i=m$ and $D_{ij}=D \neq I_m$ where $I_m$ is $m\times m$ identity matrix. This corresponds to processes that are  homogeneous meaning that the local dynamics in each node is described with the same evolution equation.    
Equations (\ref{eq1}), for $i=1, \ldots N$, can be rewritten as 
\begin{equation} \label{eq2}
\mathbf{x}(t+1) = \left( B \otimes D \right) \mathbf{x}(t) \equiv  H\mathbf{x}(t) 
\end{equation}
Assume that both matrices $B$ and $D$ are stochastic, not necessarily irreducible, and their dominant eigenvectors are $\pi$ and $\rho$ respectively. Also assume that $\pi$ and $\rho$ are normalized so that $\left\| \pi \right\| = \left\| \rho \right\|=1$. Analogous to the random walk/consensus case, here we observe 4 models, for each combination of row/column stochastic matrices $B$ and $D$.
In the rest of the paper we will use several times the following two facts. First, if $u$ is column vector and $v$ is row vector then $u \otimes v = v \otimes u = uv$; second, if $A$ is irreducible column stochastic matrix then the columns of $A^\infty$ converge to the dominant eigenvector of $A$. It is easy to see that a similar rule applies to row stochastic matrices if you take $({A^T})^\infty$.


\subsection{Random walk} \label{subsec:general-rwalk}

Assume first that model both $B$ and $D$ are column stochastic. 
%
%
The process satisfies the consistency $\left\| \mathbf{x}(t+1) \right\|=\left\| \mathbf{x}(t) \right\|$ under no additional constraints. Indeed,
\begin{eqnarray*}
\mathbf{1}_{m}^T\sum_{i=1}^{N}\mathbf{x}_i(t+1)  &=&  \sum_{i=1}^N\sum_{j=1}^Nb_{ij} \mathbf{1}_{m}^T D \mathbf{x}_j(t)   \\
& = &  \sum_{j=1}^N \mathbf{1}_{m}^T \mathbf{x}_j(t)\sum_{i=1}^N b_{ij}   = \mathbf{1}_{m}^T \sum_{j=1}^N \mathbf{x}_j(t)
\end{eqnarray*}
It corresponds to a \textit{random walk} where $\left\| \mathbf{x}(0) \right\|$ walkers move between the internal states of all the nodes. 
Note that when the number of walkers is 1,  $\left\| \mathbf{x}(0) \right\|=1$,  
the vector  $\mathbf{x} = [ \mathbf{x}_1 \ldots \mathbf{x}_N]^T$ is a probability vector, and (\ref{eq2}) describes a Markov chain. 
When both $B$ and $D$ are irreducible, the stationary solution of the random walk is 
\begin{eqnarray*} 
\mathbf{x}(t) &=& H^t \mathbf{x}(0)   \\
 & \rightarrow &  ( \pi \otimes \mathbf{1}_N^T) \otimes (\rho \otimes \mathbf{1}_m^T)  \mathbf{x}(0) \\
 & = & (\pi \otimes \rho) \otimes (\mathbf{1}_N^T \otimes \mathbf{1}_m^T) \mathbf{x}(0) \\
 & = & (\pi \otimes \rho) \left\| \mathbf{x}(0) \right\|
\end{eqnarray*}
Here, the equilibrium solution depends on both $B$ and $D$ and only on $\left\| \mathbf{x}(0) \right\|$: 
\begin{eqnarray} \label{eq-random-hom} 
\lim_{t\to \infty} {x}_i^k(t) & = &  \left\| \mathbf{x}(0) \right\| \pi_i\rho_k   
\end{eqnarray}
for all $i=1,2,\ldots,N$ and $k=1,2, \ldots, m$.  
\textit{Note that for $m=1$,  the last equation (\ref{eq-random-hom}) reduces to (\ref{eq-lin-res}).}


\subsection{Consensus} \label{subsec:general-consensus}

Assume now that both $B$ and $D$ are row stochastic. 
%
%
The model lacks consistency. When both $B$ and $D$ are irreducible, it corresponds to a \textit{consensus} where each internal variable of each node converges to the same value. Its stationary solution is
\begin{eqnarray*} 
\mathbf{x}(t) &=& H^t \mathbf{x}(0)   \\
 & \rightarrow &  ( \pi^T \otimes \mathbf{1}_N) \otimes (\rho^T \otimes \mathbf{1}_m)  \mathbf{x}(0) \\
 & = & (\pi^T \otimes \rho^T) \otimes (\mathbf{1}_N \otimes \mathbf{1}_m) \mathbf{x}(0) \\
 & = & \mathbf{1}_{Nm} (\pi^T \otimes \rho^T) \mathbf{x}(0)
\end{eqnarray*}
Here, the equilibrium solution depends on $B$ and $D$ including $\mathbf{x}(0)$: 
\begin{eqnarray} \label{eq-cons-het}  
\lim_{t\to \infty} {x}_i^k(t) & = &  (\pi^T \otimes \rho^T) \mathbf{x}(0)  
\end{eqnarray}
for all $i=1,2,\ldots,N$ and $k=1,2, \ldots, m$. 
\textit{We note again that if $m=1$,  the last equation (\ref{eq-cons-het}) reduces to (\ref{eq-lin-res}).}


\subsection{Network of Markov chains} \label{sec:network-markov-chains}

Let $B$ be row stochastic and $D$ be column stochastic. When $B$ and $D$ are irreducible, this process reaches consensus with every node having the same state (vector)
\begin{eqnarray*} 
\mathbf{x}(t) &=& H^t \mathbf{x}(0)   \\
 & \rightarrow &  ( \pi^T \otimes \mathbf{1}_N) \otimes (\rho \otimes \mathbf{1}_m^T)  \mathbf{x}(0) \\
 & = & (\mathbf{1}_N \otimes \rho) \otimes (\pi^T \otimes \mathbf{1}_m^T) \mathbf{x}(0) \\
 & = & (\pi_1 \left\| \mathbf{x}_1(0) \right\| \ldots \pi_N \left\| \mathbf{x}_N(0) \right\|)[\rho \ldots \rho]^T
\end{eqnarray*}
i.e., 
\begin{eqnarray*} 
\lim_{t\to \infty} {x}_i^k(t) & = &  (\pi_1 \left\| \mathbf{x}_1(0) \right\| \ldots \pi_N \left\| \mathbf{x}_N(0) \right\|)\rho_k   
\end{eqnarray*}
for all $i=1,2,\ldots,N$ and $k=1,2, \ldots, m$. Now, assume that $\left\| \mathbf{x}_i(0) \right\|=c, \forall i=1,\ldots,N$. Under these constraints, the model satisfies the consistency rule  $\left\| \mathbf{x}_i(t+1) \right\|=\left\| \mathbf{x}_i(t) \right\| = c$. Indeed,
\begin{eqnarray*}
\mathbf{1}_{m}^T\mathbf{x}_i(t+1)  &=&  \sum_{j=1}^Nb_{ij} \mathbf{1}_{m}^T D \mathbf{x}_j(t)   \\
& = &  \sum_{j=1}^Nb_{ij} \mathbf{1}_{m}^T \mathbf{x}_j(t)   = c \sum_{j=1}^Nb_{ij} =c
\end{eqnarray*}
since $\sum_j b_{ij}=1$ and $\mathbf{1}_m^T D = \mathbf{1}_m^T$. If $c=1$, the process corresponds to $N$ Markov chains. Since $\left\| \mathbf{x}_i(t) \right\| = 1$ at each node, (\ref{eq2}) describes a network of $N$  Markov chains. 
Each node's state corresponds to an internal (local) Markov chain, but with transition probabilities that depend on the current states of that node and on the states of the neighboring nodes. With $\left\| \mathbf{x}_i(t) \right\| = c$, only $D$ has to satisfy irreducibility in order to reach consensus: 
\begin{equation} \label{eq:homo-net-of-MC} 
\lim_{t\to \infty} {x}_i^k(t) =  c\rho_k\left\| \mathbf{b}_i \right\|=c\rho_k
\end{equation}
for all $i=1,2,\ldots,N$ and $k=1,2, \ldots, m$. Here $\left\| \mathbf{b}_i \right\|$ gives the sum of $i$-th row of $B^\infty$. Therefore, \textit{in a network of $N$ identical Markov chains, the equilibrium solution does not depend on the graph topology}. Another interesting fact is that this solution \textit{does not depend} on the irreducibility of $B$. This has several implications. First, matrix $B$ can be an arbitrary matrix. For instance, let us assume that the topology of the network is changing over time and let the matrix $B(t)$ describe the topology at time $t$. Then we have $\mathbf{x}(t) = (B(t) \otimes D)\mathbf{x}(t-1)=(B(t)B(t-1)\ldots B(1) \otimes D^t)\mathbf{x}(0)$. Since a product of two stochastic matrices is a stochastic matrix the equilibrium solution does not depend on $B$. Second, the convergence rate of the system will only depend on the second largest eigenvalue of $D$.



Now consider the case when $B$ is column stochastic and $D$ is row stochastic. When $B$ and $D$ are irreducible, this process reaches consensus within each node's state, i.e. every component of a node's state vector reaches the same value
\begin{eqnarray*} 
\mathbf{x}(t) &=& H^t \mathbf{x}(0)   \\
 & \rightarrow &  ( \pi \otimes \mathbf{1}_N^T) \otimes (\rho^T \otimes \mathbf{1}_m)  \mathbf{x}(0) \\
 & = & (\pi \otimes \mathbf{1}_m) \otimes (\mathbf{1}_N^T \otimes \rho^T) \mathbf{x}(0) \\
 & = & (\rho_1 \left\| \mathbf{y}_1(0) \right\| \ldots \rho_m \left\| \mathbf{y}_m(0) \right\|) [\pi_1\mathbf{1}_m \ldots \pi_N\mathbf{1}_m]^T
\end{eqnarray*}
i.e.,
\begin{eqnarray*} 
\lim_{t\to \infty} {x}_i^k(t) & = &  (\rho_1 \left\| \mathbf{y}_1(0) \right\| \ldots \rho_m \left\| \mathbf{y}_m(0) \right\|)\pi_i   
\end{eqnarray*}
for all $i=1,2,\ldots,N$ and $k=1,2, \ldots, m$. Here, $\mathbf{y}_k(t) = [ x_1^k(t) \ldots x_N^k(t)]^T$ is a column vector of length $N$ containing the $k$-th component of each node's state.
 
Now, assume that $\left\| \mathbf{y}_i(0) \right\|=c, \forall i=1,\ldots,m$. Under these constraints, the model satisfies the consistency rule $\left\| \mathbf{y}_i(t+1) \right\|=\left\| \mathbf{y}_i(t) \right\| = c$. Indeed,
\begin{eqnarray*}
\sum_{i=1}^N{\mathbf{x}_i(t+1)}  &=&  \sum_{i=1}^N\sum_{j=1}^N b_{ij} D \mathbf{x}_j(t)   \\
& = &  D\sum_{j=1}^N \mathbf{x}_j(t)\sum_{i=1}^N b_{ij}   = Dc\mathbf{1}_m = c\mathbf{1}_m,
\end{eqnarray*}
since $\sum_i b_{ij}=1$ and $D\mathbf{1}_m = \mathbf{1}_m$.
If $c=1$, the process corresponds to $m$ different Markov chains. In this case, the model describes a consensus between $m$ Markov chains, each chain having $N$ states. With $\left\| \mathbf{y}_i(t) \right\| = c$, only $B$ has to satisfy irreducibility in order to reach consensus:
\begin{equation*} 
\lim_{t\to \infty} {x}_i^k(t) = c\pi_i\left\| \mathbf{d}_k \right\|=c\pi_i   
\end{equation*}
for all $i=1,2,\ldots,N$ and $k=1,2, \ldots, m$. Here $\left\| \mathbf{d}_k \right\|$ gives the sum of $k$-th row of $D^\infty$. Therefore, \textit{in a network of $N$ nodes, in which each node is described with $m$ dimensional state vector such that every component of a node's state vector corresponds to a Markov chain, the equilibrium solution depends only on the graph topology.} Here the solution does not depend on the irreducibility of $D$ which has implications analogous to the previous process.

In fact, the two models described in this subsection are related to each other. Indeed, let $\mathbf{y}_k(t) = [ x_1^k(t) \ldots x_N^k(t)]^T$ be a column vector of length $N$ containing the $k$-th component of each node's state and let $\mathbf{y} = [ \mathbf{y}_1 \ldots \mathbf{y}_m]^T$ be a column vector of length $mN$. Equation (\ref{eq2}) can be rewritten as:
\begin{equation*}
 \mathbf{y}(t+1) = \left( D \otimes B \right) \mathbf{y}(t) 
\end{equation*}

\textit{Remark 1.}
Assume that $D$ is double stochastic. We now have only two models depending on $B$ with stationary solutions $x_i^k=\left\|\mathbf{x}(0)\right\|/m$ and $x_i^k=\pi_i \left\|\mathbf{x}(0)\right\|/m$ when $B$ is row and column stochastic respectively. On the other hand, when $B$ is double stochastic, the stationary solutions are $x_i^k=\left\|\mathbf{x}(0)\right\|/N$ and $x_i^k=\rho_k \left\|\mathbf{x}(0)\right\|/N$ when $D$ is row and column stochastic respectively. When both $B$ and $D$ are double stochastic, all four models collapse into one model with stationary solution $x_i^k=\left\|\mathbf{x}(0)\right\|/(Nm)$.

\textit{Remark 2.}
For $D=I_m$, equations (\ref{eq1}), for $i=1, \ldots N$, can be rewritten as $m$ equations $ \mathbf{y}_k(t+1) =  B\mathbf{y}_k(t)$, for all $k=1,\ldots, m$.
Therefore, when $D=I_m$, the model (\ref{eq1}) decouples into $m$ models of (\ref{eq-linear}). Similarly, when $B=I_N$, the model (\ref{eq1}) decouples into $N$ models of (\ref{eq-linear}).

\section{Heterogeneous linear processes} \label{sec:hetero}

In this section we focus on the processes which we call heterogeneous,  where the local dynamics differs for each node in the network. Equation (\ref{eq1}) can be rewritten in more compact form using the following extension of  Kronecker product of matrices $B$ and $D_{ij}$. 
Define    $H = B \otimes \{D_{ij} \} \equiv [H_{ij}]$, where $H_{ij} = b_{ij}D_{ij}$.  Then Eq.~(\ref{eq1}) becomes:  
\begin{equation}
\label{eq-heter}
\mathbf{x}(t+1) =  B \otimes \{D_{ij} \}\mathbf{x}(t) =   H \mathbf{x}(t),  
\end{equation}
or, in the equivalent form,  the Eq.~(\ref{eq1}) can be rewritten as: 
$$
\mathbf{y}(t+1) =   H \mathbf{y}(t), 
$$ 
where 
$$
\mathbf{y} = [\underbrace{y_1 y_2 \ldots y_{m_1}}_{\mathbf{x}_1} \underbrace{y_{m_1+1} y_{m_1 + 2} \ldots  y_{m_1 + m_2}}_{\mathbf{x}_2} \ldots y_s]^T
$$
is  a column vector of length $s=m_1 +\ldots +  m_N$ and $H$ is  a  $s \times s$ matrix.

\subsection{Random walk} 

Using similar arguments as above, it can be shown that when $B$ is column stochastic and each column of $D_{ij}$ sums up to 1, the model  satisfies the consistency $\left\| \mathbf{x}(t+1) \right\|=\left\| \mathbf{x}(t) \right\|$ under no additional constraints. Moreover, $H$ is column stochastic matrix. To see this, note that $h_{rc}$ has the value $b_{ij}d_{ij}^{kl}$ with $r=\sum_{t=1}^{i-1}m_t+k$ and $c=\sum_{t=1}^{j-1}m_t+l$. Then, note that $r$ does not depend on $j$ and $l$. Therefore,
\begin{equation*}
 \sum_{r=1}^{s}h_{rc} = \sum_{i,k}b_{ij}d_{ij}^{kl}  = \sum_{i}b_{ij}\sum_{k}d_{ij}^{kl} = \sum_{i}b_{ij} = 1
\end{equation*}
 
Assuming  that $H$ is irreducible matrix, the multiplicity of its dominant eigenvalue is 1 and asymptotic behavior of the Eq.~(\ref{eq-heter})  is determined by the eigenvector corresponding to 1, i.e. 
\begin{eqnarray*} 
\lim_{t\to \infty} {y}_i(t) & = &  \left\| \mathbf{x}(0) \right\| \gamma_i    
\end{eqnarray*}
for all $i=1,2, \ldots, s$, where $\gamma= [\gamma_1, \gamma_2, \ldots \gamma_{s}]^T$ is the  dominant eigenvector of $H$.  
\textit{Note that for $D_{ij}=D$,  the last equation reduces to (\ref{eq-random-hom}).}
In fact, this process corresponds to the 2-step random walk discussed in example 1. Therefore, 2-step random walk can always be seen as a regular random walk with one transition matrix $H = [b_{ij}d_{ij}^{kl}]$, thus treating each city as a seperate country. However, knowing that there are patterns of repetition in the probabilities, we can group cities into countries, i.e. we can decompose matrix $H$ into matrices $B$ and $D_{ij}$. The advantage in doing this is that, if we were interested in country dynamics only, instead of solving the full system given by Eq.~(\ref{eq1}), we can solve a smaller model with $N$ states. Indeed, 
 \begin{equation}
\label{eq:small-model}
 \left\|\mathbf{x}_i(t+1)\right\| =   \sum_{j=1}^n b_{ij}\mathbf{1}_{m_i}^TD_{ij}\mathbf{x}_j(t)= \sum_{j=1}^n b_{ij}\left\| \mathbf{x}_j(t) \right\|
\end{equation}
Note that we can not obtain the city dynamics from country dynamics. However, if choosing a city (the second step) didn't depend on the current city, but rather only on the current country, we can further decompose matrices  $D_{ij}$ into $\mathbf{d}_{ij}\mathbf{1}_{m_j}^T$ where $\mathbf{d}_{ij}$ is a column stochastic vector of length $m_i$ describing the transition probabilities from country $j$ to the $m_i$ cities in country $i$. Then Eq.~(\ref{eq1}) gets the following form:
\begin{equation} \label{eq:smaller-model}
 \mathbf{x}_i(t+1) = \sum_{j=1}^n b_{ij}\mathbf{d}_{ij}\mathbf{1}_{m_j}^T\mathbf{x}_j(t) = \sum_{j=1}^n b_{ij}\mathbf{d}_{ij}\left\| \mathbf{x}_j(t) \right\|
\end{equation}
Here, there is a clear connection between country dynamics $\left\| \mathbf{x}_i\right\|$ and city dynamics $\mathbf{x}_i$. Therefore, if we have the country dynamics, we can obtain the city dynamics. Note that this equation is local in terms of country $i$, i.e. in order to obtain city dynamics (from country dynamics) for country $i$, the decomposition of $D_{ij}$ has to apply only for countries that point to country $i$. Also note that the decomposition of $D_{ij}$ for all $i$ is satisfied by definition for each country $j$ that has only one city, since the size of $D_{ij}$ will be $m_i\times1$.

\subsection{Consensus} 
Assume now that $B$ is row stochastic and  each row of $D_{ij}$ sums up to 1. Then using similar arguments as above, it can be shown that $H$ is row stochastic as well. Further, assuming  that $H$ is irreducible matrix, one can show
\begin{eqnarray*} 
\lim_{t\to \infty} {y}_i(t) & = & \sum_i \gamma_i y_i(0)      
\end{eqnarray*}
for all $i=1,2, \ldots, s$.
\textit{Note again that for $D_{ij}=D$,  the last equation reduces to (\ref{eq-cons-het}).}
This proccess can be seen as a regular consensus with one row stochastic matrix $H = [b_{ij}d_{ij}^{kl}]$. Similarly to the random walk process, we can take advantage of the decomposition of $H$ by approximating the stationary solution. To do this, we solve the smaller system (\ref{eq:small-model}) to calculate the sum of the weights for each external node $k$, i.e.  $\Gamma_k =\left\|\mathbf{x}_k(\infty)\right\|=\sum_{l\in k}\gamma_l$ for each $k$. Then, assuming that the internal weights are equal, we can obtain the approximation as
\begin{equation*} 
\lim_{t\to \infty} {y}_i(t)=\sum_{k}\frac{\Gamma_k}{|k|}\sum_{l\in k} y_l(0)
\end{equation*}
where $|k|$ denotes the number of internal nodes in node $k$. Furthermore, the smaller the fluctuations of $y_l(0)$ within a node, the better the approximation. In fact, if $y_i(0)=y_j(0)$ holds for all $i,j\in k$ and for all $k$, the approximation becomes exact regardless of whether the actual internal weights are equal.
Note that, when $D_{ij}=D$, this process becomes identical to the opinion formation model discussed in example 3. Therefore, if $H$ is irreducible, the opinion formation model will reach consensus where there is a single opinion $\sum_i \gamma_i y_i(0)$ valid for every topic and every person in the network. However, since the model discussed in example 3 is also homogeneous, we don't have to find the dominant eigenvector of $H$, but the eigenvectors of smaller matrices $B$ and $D$ as shown in Section \ref{subsec:general-consensus}.

\subsection{Network of Markov chains}

Finally we consider the case when $B$ is row stochastic and each column of $D_{ij}$ sums up to 1.  
Then in general, the matrix $H$ is not stochastic, however 1 is its dominant eigenvalue. This follows from the fact that $H$ is derived from a convex combination of stochastic matrices.  
The multiplicity of this eigenvalue is tied to the structure of the underlying graph.    
If $H$ is irreducible matrix, then the multiplicity of its dominant eigenvalue is 1 and asymptotic behavior of the Eq.~(\ref{eq-heter})  is determined by the eigenvector corresponding to 1. 
It can be also shown that in this case the consistency rule is satisfied at local level: at each node, ($\left\| \mathbf{x_i}(t+1) \right\|=\left\| \mathbf{x_i}(t) \right\|$), and therefore, in this case,  \textit{the model corresponds to a network of $N$ different Markov chains}.  

Let $\gamma= [\gamma_1, \gamma_2, \ldots \gamma_{s}]^T$ be the eigenvector of $H$ corresponding to the eigenvalue 1. Let $\alpha_i$ be a stationary distribution of the Markov chain at node $i$.
Then using similar arguments as above, it can be shown that
$$
\alpha_i = [\gamma_{a+1} \ldots \gamma_{a+m_i}]^T.
$$
where $a=\sum_{j=1}^{i-1}m_j$. In the special case when $D_{ij}=D$, we have  
$$
\alpha_i = \alpha = [\gamma_1 \ldots \gamma_m]^T
$$  
for all $i$, where  $\gamma_k = \rho_k$ for all $k$ and 
$\rho=[\rho_1 \ldots \rho_m]^T$ is a dominant eigenvector of $D$.
Therefore, all Markov chains reach consensus state, as discused in Section \ref{sec:network-markov-chains}. \textit{Note that for $D_{ij}=D$,  the last equation reduces to (\ref{eq:homo-net-of-MC}) for $c=1$.}
In fact, this process corresponds to the weather model discussed in example 2.

\begin{figure*}[htb]
\begin{center}
\includegraphics[scale=0.8]{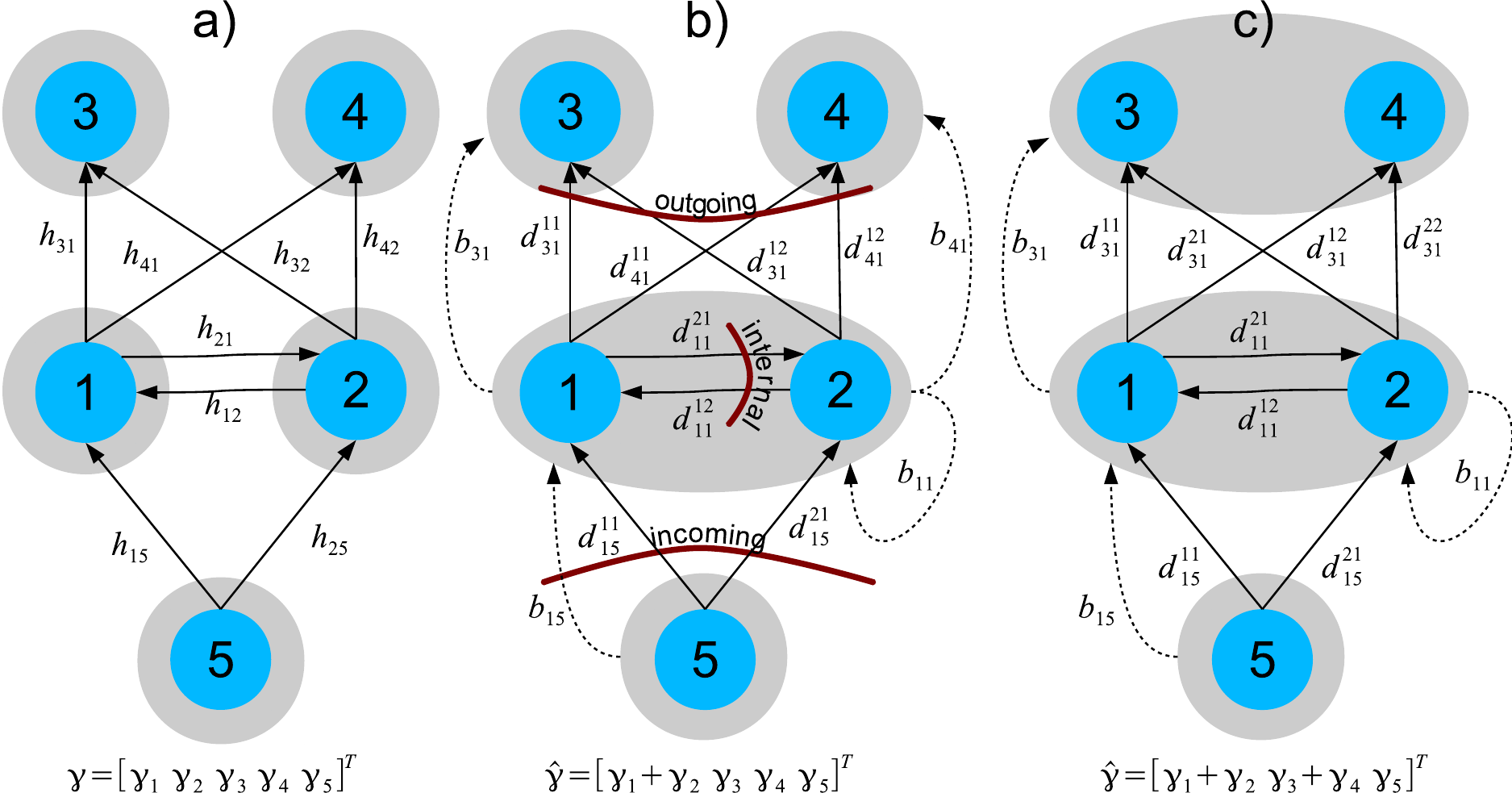}
\end{center}
\caption{Heterogeneous random walk model with lumping. Nodes correspond to cities where each city $i$ belongs to its own country $i$ unless otherwise noted. The stationary solution vector of the corresponding lumped system is given on the bottom of each example. a) Regular random walk with transition matrix $H$ equivalent to heterogeneous random walk where each city belongs to its own country, making $B=H$ and $D_{ij}=1$. Solid lines depict the transition probabilities. b) Cities $1$ and $2$ share the same country, i.e. they are the first and the second city in country $1$ respectively. c) Cities $3$ and $4$ also share a country, i.e. they are the first and the second city in country $3$ respectively. b) and c) Dashed lines depict the country transition probabilities and solid lines depict the city transition probabilities.}
\label{fig:inverse}
\end{figure*}

\section{Network hierarchy} \label{sec-net-hier} 

Complex systems including networks often exhibit hierarchical organization in which the network self-organizes into modules that further subdivide into modules of modules, and so forth over multiple scales \cite{clauset}. 
This way, hierarchical systems evolve much more rapidly from elementary constituents than non-hierarchical systems with the same number of elements \cite{simonh}. Very often these hierarchies exhibit the concept of ``near-decomposability''. Put as simply as possible, it is the degree to which the behavior of a system at any one level is free of the interactions on a lower level and the degree which its interactions are irrelevant to the higher levels of the system.
%
In many cases the groups (modules) are found to correspond to known functional units, such as ecological niches in food webs, modules in biochemical networks (protein interaction networks, metabolic networks or genetic regulatory networks) or communities in social networks \cite{clauset,clauset2,ravasz,guimera,lagomarsino}.

Typically, a network hierarchy is a product of dynamical processes that govern the evolution of the network. We now argue that the approach for studying linear processes on networks developed in this paper can also be extended to study network hierarchy.  
In general, two problems related to network hierarchy can be posed. The first problem consists of developing a general framework to study network hierarchy taking into account the processes on networks. The second problem aims at decomposing a given graph -- the matrix $H$ -- into subgraphs (groups) that are described by matrices $B$ and $D_{ij}$.  Clearly this problem is very important having in mind how huge real networks are.  

To study the first problem, we note that the model  (\ref{eq-heter}) describes a homogeneous process occurring on 2-level hierarchical network described through the matrices $B$ and $D_{ij}$. 
What is interesting is that these processes can be generalized to fit an arbitrary hierarchy of interactions of modules. More specifically, we introduce the hierarchical linear processes:
\begin{eqnarray} \label{eq-hier}
\mathbf{x}(t+1) = \left( \ldots \left( B \otimes \{ D^1_{ij} \} \right) \ldots \otimes \{ D^h_{ij} \}  \right) \mathbf{x}(t), 
\end{eqnarray}
where $H_1 = B \otimes \{D^1_{ij}\}$, $H_2 = H_1 \otimes \{D^2_{ij}\}$, so on and
$H_h = H_{h-1}  \otimes \{D^h_{ij}\}$, 
describing interactions on $h$-level hierarchical complex systems (networks). Here, $\mathbf{x}$ is a column vector of length $N$, the number of elementary constituents in the system, and $H_i$ is an $N_i \times N_i$ matrix describing the interactions occurring between $N_i$ modules in the $i$-th level of the hierarchy. Assume that matrices $H_i$ for all $i$ are stochastic, irreducible and their dominant eigenvectors are $\pi^i$ respectively. 
In the special case when the process (\ref{eq-hier}) is homogeneous
($D_{ij}^k =D^k$), it creates decomposability in the asymptotic behavior of the system, where the solution can be expressed as a function of global parameters (network topology) and local parameters (local dynamics).   
Due to the homogeneity of the system, the solution of (\ref{eq-hier}) can be expressed as a function of the eigenvectors $\pi^i$.
%

%


The second problem, namely the problem of decomposing the matrix $H$ into $B$ and $D_{ij}$, will be addressed only by considering an example. 
%
The decomposition of $H$ has several advantages. In the random walk case, the stationary solution is one of the most used centrality measures in networks and its well known variant, the PageRank algorithm \cite{pagerank}, lies at the heart of Google's search technology. Here, the decomposition can be used to obtain a high-level view of stationary dynamics by lumping nodes into super-nodes. This reduces the size of the system, and thus, the time it takes to compute the solution. In the consensus case, on the other hand, we have shown that the high-level stationary solution can be used to obtain approximation for the stationary consensus value. The benefits we get from decomposing $H$ are clear, but we still have not discussed the actual problem of decomposing a given matrix $H$ into $B$ and $D_{ij}$. In order to do so, $H$ has to satisfy some properties. In order to understand these properties, we present a toy graph where we discuss the random walk case; we omit the discussion for the consensus case because of the well known duality between the two processes.

The graph is shown in Fig. \ref{fig:inverse}  and consists of five nodes with directed edges denoting transition probabilities. For consistency, we will refer to external nodes as countries and internal nodes as cities, and use the same notations from previously, i.e. $h_{ij}$ denotes the transition probability from city $j$ to city $i$, $b_{ij}$ the transition probability from country $j$ to country $i$, and $D_{ij}^{kl}$ the transition probability from the $l$-th city in country $j$ to the $k$-th city in country $i$. We start with Fig. \ref{fig:inverse}a by showing a regular random walk with transition matrix $H$. We can think of this model as 2-step random walk where each city is in its own country, with $1\times1$ matrix $D_{ij}=1$ and $B=H$. Now assume that city $1$ and city $2$ belong to the same country, i.e. they are the first and the second city in country $1$ respectively, and every other city belongs to its own country (see Fig. \ref{fig:inverse}b). This means that the transition probabilities from cities $1$ and $2$ are influenced by a common country decision factor $b_{j1}$. To see this further, we start by describing the outgoing links from country $1$, i.e.
\begin{equation*}
\begin{array}{l}
\displaystyle h_{31}=b_{31}d_{31}^{11} \mbox{ and } h_{32}=b_{31}d_{31}^{12}\\
\displaystyle h_{41}=b_{41}d_{41}^{11} \mbox{ and } h_{42}=b_{41}d_{41}^{12}
\end{array} 
\end{equation*}
where $D_{31}$ and $D_{41}$ are $1\times2$ column stochatic matrices. This means that $d_{31}^{11}=d_{31}^{12}=d_{41}^{11}=d_{41}^{12}=1$. Therefore we have
\begin{equation}
\begin{array}{l}
\displaystyle h_{31}=h_{32}=b_{31}\\
\displaystyle h_{41}=h_{42}=b_{41}
\end{array} 
\label{eq:outgoing}
\end{equation}
This tells us that cities $1$ and $2$ have the same distribution of transition probabilities. Furthermore, let us look at the internal links in country $1$
\begin{equation*}
\begin{array}{l}
\displaystyle h_{11}=b_{11}d_{11}^{11} \mbox{ and } h_{21}=b_{11}d_{11}^{21}\\
h_{12}=b_{11}d_{11}^{12} \mbox{ and } h_{22}=b_{11}d_{11}^{22}
\end{array} 
\end{equation*}
with matrix $D_{11}$ describing the local transition probabilities. Summing $h_{11}$ and $h_{21}$ and also, $h_{12}$ and $h_{22}$ together with the fact that $D_{11}$ is column stochastic, we have
\begin{equation} 
\label{eq:internal}
h_{11}+h_{21}=h_{12}+h_{22}=b_{11}
\end{equation}
This means that the probability of staying within country borders is the same regardless of whether the walker is currently at city $1$ or $2$. To sum up, Eq.~(\ref{eq:outgoing}) tells us that cities that share country $j$, share also the same outgoing probability distribution (same set of weighted outgoing links) to outside cities, given by $b_{kj}$, for all cities $k\notin j$ (assuming that each city $k$ is in its own country $k$). Eq.~(\ref{eq:internal}), on the other hand, tells us that cities that share country $j$ also have the same aggregated probability of staying within country $j$, given by $b_{jj}$. Therefore, from the country point of view, it is statistically irrelevant to know the actual city the walker is visiting. This conclusion is supported by (\ref{eq:small-model}) where nodes that share the same outgoing weighted links can be lumped in order to descrease the number of states in the system. Actually, this is a well known technique used in PageRank computation where all dangling nodes (nodes with no outlinks) are lumped in one node \cite{pagerank-dangling}. This is done by solving a smaller system of size $n+1$ (see system (\ref{eq:small-model})), where $n$ is the number of non-dangling nodes and then its solution is used for one iteration of system (\ref{eq:smaller-model}) in order to get the ranking for the dangling nodes as well. Note that we can get the ranking for the dangling nodes since all other nodes are actually external nodes with size $1$, thus the decomposition of $D_{ij}$ is satisfied by definition. What is left is to look into the incoming links for country $1$, i.e. the outgoing links of city $5$
\begin{equation*}
\begin{array}{l}
\displaystyle h_{15}=b_{15}d_{15}^{11} \mbox{ and } h_{25}=b_{15}d_{15}^{21}
\end{array} 
\end{equation*}
By summing $h_{15}$ and $h_{25}$ and using the fact that $D_{15}$ is column stochastic we get the transition probabilities from city $5$ to country $1$ as
\begin{equation*}
\begin{array}{l}
\displaystyle h_{15}+h_{25}=b_{15} \\
\displaystyle d_{15}^{11}=\frac{h_{15}}{h_{15}+h_{25}} \\
\displaystyle d_{15}^{21}=\frac{h_{25}}{h_{15}+h_{25}}
\end{array} 
\end{equation*}
This tells us that if lump cities into countries, we can obtain the new transitional probabilities by using simple normalization. The properties that cities should posses in order to lump them into countries are pretty strict. This is because we are treating all the other nodes as countries with a single city. However, when we have several countries with more than one city, these properties are relaxed. To show this, assume also that city $3$ and $4$ are sharing the same country, i.e. they are the first and the second city in country $3$ respectively (see Fig. \ref{fig:inverse}c). Then we have
\begin{equation*}
\begin{array}{l}
\displaystyle h_{31}=b_{31}d_{31}^{11} \mbox{ and } h_{41}=b_{31}d_{31}^{21}\\
\displaystyle h_{32}=b_{31}d_{31}^{12} \mbox{ and } h_{42}=b_{31}d_{31}^{22}
\end{array} 
\end{equation*}
Summing $h_{31}$ and $h_{41}$ and also, $h_{32}$ and $h_{42}$ together with the fact that $D_{31}$ is column stochastic, we have
\begin{equation} 
\label{eq:relaxed}
h_{31}+h_{41}=h_{32}+h_{42}=b_{31}
\end{equation}
We can see that the property for lumping nodes is relaxed. Eq.~(\ref{eq:relaxed}) tells us that cities that share country $j$, share also the same aggregated probability distribution for outside countries, given by $b_{ij}$, for all countries $i$.

To conclude this section, we stress that heterogeneous random walk processes satisfy an important property that allows us to lump nodes without loosing much information about the system (see (\ref{eq:small-model})). In fact, if the super-nodes are not sharing any links, no information will be lost after lumping. That is, if only regular nodes point to these super-nodes, we can easily restore the internal dynamics after computing the high-level dynamics (see (\ref{eq:smaller-model})). We have also shown the properties that nodes must have in order to share the same super-node. Given a matrix $H$, the problem of finding the right partition of nodes, i.e. decomposing $H$ into $B$ and $D_{ij}$ is clearly important. We believe this problem belongs to the class of NP problems, however the study of this problem is beyond the scope of this paper and will be discussed in a forthcoming paper.

\section{Conclusions}  \label{sec:concl}

In conclusions, we have introduced a broad class of analytically solvable processes on networks. Both homogeneous and heterogeneous models are analytically solvable, although for homogeneous models, it was shown that the analytical solution is explicit function of two sets of parameters: one being the parameters of the network topology and the other being the parameters of the local (node) dynamics. Finally, we have shown that suggested framework for analysis of linear processes on networks can be taken as advantage and can be used to model and/or to understand  interactions on hierarchical complex systems.


\end{document}